# Holographic imaging of the complex charge density wave order parameter

Árpád Pásztor, Alessandro Scarfato, Marcello Spera, Céline Barreteau, Enrico Giannini, and Christoph Renner*

*DQMP, Université de Genève, 24 quai Ernest Ansermet, CH-1211 Geneva 4, Switzerland.*

**The charge density wave (CDW) in solids is a collective ground state combining lattice distortions and charge ordering. It is defined by a complex order parameter with an amplitude and a phase. The amplitude and wavelength of the charge modulation are readily accessible to experiment. However, accurate measurements of the corresponding phase are significantly more challenging. Here we combine reciprocal and real space information to map the full complex order parameter based on topographic scanning tunneling microscopy (STM) images. Our technique overcomes limitations of earlier Fourier space based techniques to achieve distinct amplitude and phase images with high spatial resolution. Applying this analysis to transition metal dichalcogenides provides striking evidence that their CDWs consist of three individual charge modulations whose ordering vectors are connected by the fundamental rotational symmetry of the crystalline lattice. Spatial variations in the relative phases of these three modulations account for the different contrasts often observed in STM topographic images. Phase images further reveal topological defects and discommensurations, a singularity predicted by theory for a nearly commensurate CDW. Such precise real space mapping of the complex order parameter provides a powerful tool for a deeper understanding of the CDW ground state whose formation mechanisms remain largely unclear.**

The spatially averaged intensity of the CDW order parameter is usually accessed by scattering techniques sensitive to the local lattice distortions, or electron spectroscopy and transport measurements sensitive to changes in the band structure due to the opening of the CDW gap. Detecting the phase has been traditionally limited to dynamic experiments (for good reviews see e.g. refs 1, 2). More recently, different strategies have been followed to access phase information from spectroscopic[3, 4] and topographic[5] local probe tunneling experiments as well as from the periodic distortion of the atomic lattice measured by transmission electron microscopy[6, 7]. However, extracting phase information from reciprocal-space data tends to suffer from the common problem of phase wrapping and fundamental numerical difficulties of finite sampling (see Supplementary Information Section 3). Unwrapping the phase in these cases leads to singularities that can easily be mistaken for physical phase variations that actually do not exist and it can also obscure real features, for instance domain walls in 1T-$Cu_xTiSe_2$. Combining reciprocal and real space scanning tunneling microscopy (STM) data to extract the full complex CDW order parameter enables to overcome these limitations.

The CDW state is characterized by a real-space periodic modulation of the charge density[8]:

$$\delta\rho(\boldsymbol{r}) = \rho(\boldsymbol{r}) - \rho_0 = A \cdot cos(\boldsymbol{q} \cdot \boldsymbol{r} + \varphi) = \Re(\Psi e^{i\boldsymbol{q} \cdot \boldsymbol{r}}) \qquad (1)$$

where $\rho_0$ is the constant background charge density in the metallic state, **q** is the ordering vector and $A$ is the modulation amplitude which is proportional to the energy gap in the quasi-particle spectrum due to the formation of the CDW phase. The order parameter of the CDW state is a complex quantity: $\Psi = A \cdot e^{i\varphi}$. In general, one expects its amplitude and phase to spatially vary ($A = A(\mathbf{r})$ and $\varphi = \varphi(\mathbf{r})$), leading to collective excitations (amplitudons and phasons, respectively) and defects (domain walls, vortices, discommensurations, etc…) in the CDW condensate.

In this study, we focus on the complex CDW order parameter in three transition metal dichalcogenides (TMDCs) $MSe_2$. They are layered materials where each slab is composed of a triangular metal layer M ($Cu_xTi$, Nb and V in the present study) sandwiched between two triangular selenium sheets. The layer stacking is maintained by weak van der Waals (vdW) forces allowing facile cleaving between the slabs exposing a triangular Se layer to the surface. In 1T-$Cu_xTiSe_2$, copper atoms intercalate on the octahedral site in the vdW gap between the $TiSe_2$ slabs[9]. 1T-$Cu_xTiSe_2$ and 1T-$VSe_2$ host a short-range $2a \times 2a$ and a commensurate $4a \times 4a$ CDW within the basal plane[9-12], respectively, while 2H-$NbSe_2$ has been shown to develop a slightly incommensurate $(3 + \delta)a \times (3 + \delta)a$ CDW modulation[13], with $\delta \approx 0.03 - 0.08$ depending on temperature.

All three compounds studied here have a crystal structure with a three-fold symmetry in the *ab*-plane, regardless of their specific polytype (1T or 2H). The crystal symmetry implies that if there exists an ordering vector $\mathbf{q}_1$ for which the formation of a CDW is favourable in the

*ab*-plane, there are two other momentum vectors $\mathbf{q}_2$ and $\mathbf{q}_3$ for which the CDW formation is equally favourable. These three q-vectors are connected by a three-fold $2\pi/3$ rotation symmetry and are often referred to as the *three components* of the *tri-directional* CDW. The total charge density modulation in space can be expressed as the sum of these three charge density modulations

$$\delta\rho(\mathbf{r}) = \sum_{n=1}^{3} A_n(\mathbf{r})\cos(\mathbf{q}_n \cdot \mathbf{r} + \varphi_n(\mathbf{r})) = \sum_{n=1}^{3} \Re\left(\Psi_n e^{i\mathbf{q}_n \cdot \mathbf{r}}\right) \qquad (2)$$

where $\Psi_n = A_n(\mathbf{r})e^{i\varphi_n(\mathbf{r})}$ is the order parameter of the n-th direction and $\mathbf{q}_n$ and $\varphi_n$ are the corresponding ordering vectors and phase shifts. Hereafter and following the proposition of McMillan[14], we refer to the CDW modulations in the three $\mathbf{q}_n$ directions as *the three coexisting CDWs*, a denomination fully justified by our experimental findings in the following sections.

A rapid survey of the literature reveals that the CDW contrast in a given material varies significantly among published STM images, even in the absence of any structural defects. While in several cases this can be explained with changing experimental conditions, either extrinsic, e.g. a change in the tip state, or intrinsic, e.g. bias dependence, we find compelling evidence of contrast variations which are real CDW features and not the consequence of particular tunneling conditions. In Fig. 1**a**, **c-d** we highlight two adjacent regions with a different contrast in the same STM micrograph. As we show in the following, these two contrasts can be reproduced straightforwardly by tuning the relative phases of the three CDWs discussed above. To quantify the changing configurations, we introduce a dephasing parameter $\Theta$ defined as the sum of the three individual CDW phases $\varphi_n$ modulo $2\pi$:

$$\Theta = \left(\left(\sum_{n=1}^{3} \varphi_n\right) \mod 2\pi\right). \qquad (3)$$

This dephasing parameter uniquely determines the general appearance of the CDW pattern. Each $\Theta$ value corresponds to a particular CDW imaging contrast, regardless of how individual phase shifts $\varphi_n$ are distributed among the three $\Psi_n$. A more detailed description of the dephasing parameter is given in Supplementary Information Section 1.

In order to determine the phase $\varphi_n$ of each $\Psi_n$ required to compute the dephasing parameter, we developed a procedure based on the local $\mathbf{q}_n$-specific fitting of the real space CDW modulation measured by STM. We locally describe each CDW by a two-dimensional plane wave function $f_n(\mathbf{r}) = A_n(\mathbf{r})\cos\left(\frac{2\pi}{\lambda_n}\hat{\mathbf{q}}_n \cdot \mathbf{r} + \varphi_n(\mathbf{r})\right)$, where $\hat{\mathbf{q}}_n$ is a unit vector in the direction of the n-th ordering vector. We consider a model where we allow the amplitude ($A_n(\mathbf{r})$) and the phase ($\varphi_n(\mathbf{r})$) for each direction ($n = 1,2,3$) to vary spatially. The result is a complete local characterization of the CDW modulation.

To illustrate the fitting procedure described in details in Supplementary Information Section 2, we apply it to a large scale STM image of an in-situ cleaved 1T-VSe$_2$ single crystal measured at 40 K (Fig. 1). The atomic and the $4a \times 4a$ CDW modulations are clearly resolved in the micrograph (Fig. 1**a**). The red and blue squares, magnified in Fig. 1**c** and **d**, highlight two adjacent regions with different CDW contrasts. This image was acquired from top to bottom, with the fast scan direction running horizontally. The two highlighted regions are thus connected through continuous scan lines and therefore imaged with exactly the same tip at the same set point. Consequently, the distinct appearances of the CDW in these two regions cannot be imputed to changing tunneling conditions. Magnifying the same two regions of the Fourier-filtered image of Fig. 1**a** clearly shows the CDW origin of the changing contrast (Fig. 1**e** and **f**). Indeed, applying our fitting procedure allows to perfectly reproduce the experimental CDW contrast (Fig. 1**g** and **h**) characterized by two distinct dephasing parameters $\Theta$ displayed in Fig. 1**b**. Running the fitting procedure on a dense grid of slightly overlapping small windows spanning the entire field of view of Fig. 1**a** allows to reconstruct the complete spatial structure of the dephasing parameter (Fig. 1**b**). We find that the changing contrast over the entire scan area is fully reproduced by a specific local dephasing parameter with a remarkable match between the experimental and the calculated contrasts as exemplified in Fig. 1**e**-**h**.

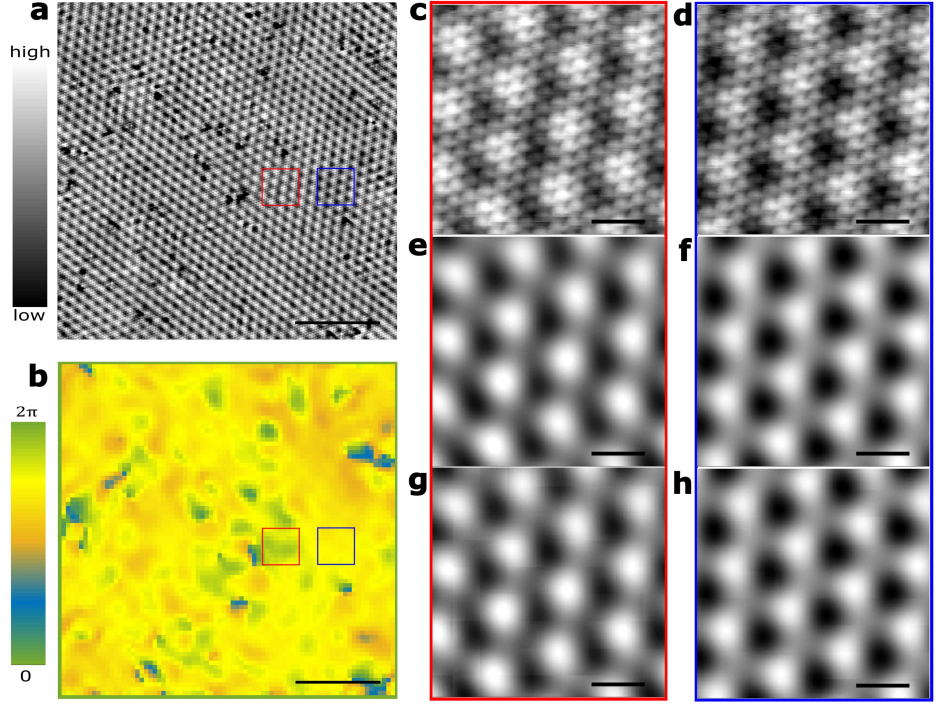

**Fig. 1 | Real space CDW pattern and the dephasing parameter. a,** High resolution STM micrograph ($V_{bias}$=-100 mV, $I_t$=1 nA) of an in-situ cleaved 1T-VSe$_2$ surface at 40 K. The atomic lattice and the CDW modulation are clearly resolved. **b,** Spatial variation of the dephasing parameter as determined by the fitting procedure. The areas marked by red and blue squares have distinct dephasing parameters corresponding to the different appearances of the CDW. **c,** and **d,** magnified areas corresponding to the red and blue squares in panel **a**, respectively. They are connected by single scan lines, the different contrasts can thus not be blamed on different tip state configurations. **e,** and **f,** are zooms into the large Fourier filtered image of the CDWs at the same location as shown in **c** and **d**, respectively. **g,** and **h,** are reconstructed images from the mosaics of the fitting procedure, in excellent agreement with **e** and **f**. The topographic images are all plotted with the exact same color scale. Scale bars: 10 nm in **a** and **b** and 1 nm in **c-h**.

So far, we have established the ability of our model to account for the variety of CDW contrasts observed in STM micrographs of a given material using a single well-defined dephasing parameter. We now advance a step further in the quantitative analysis to discuss the individual amplitudes and phases of each of the three CDWs as a function of position. The spatially resolved amplitudes $A_n(\mathbf{r})$ and the corresponding local phases $\varphi_n(\mathbf{r})$ for each $\mathbf{q}_n$, over the same field of view as Fig. 1a, are shown in Fig. 2a-c and in Fig. 2d-f, respectively. A remarkable aspect of this dataset is the absence of any obvious spatial correlation between the amplitude and phase images of the different CDWs. Each map displays its own features, meaning the amplitude and the phase of each CDW can vary at some location, even abruptly, regardless of the other two CDWs whose amplitudes and phases can remain constant and featureless at that same position. This clearly shows that these three CDWs develop their individual local complex order parameters, albeit being indistinguishable in the sense that they are connected by symmetry. We find a very similar behaviour in 1T-$Cu_{0.02}TiSe_2$ and 2H-$NbSe_2$ (see Suppl. Fig. 5 and 6), supporting the idea that the total charge density modulation in these TMDCs consists of three coexisting CDWs. Note that although we find three rather independent $\Psi_n$ in the examples discussed here, interactions mediated by the atomic lattice are possible. For example, pinning by crystalline defects or impurities can distort the charge modulations[15, 16] and introduce correlations between the three CDWs.

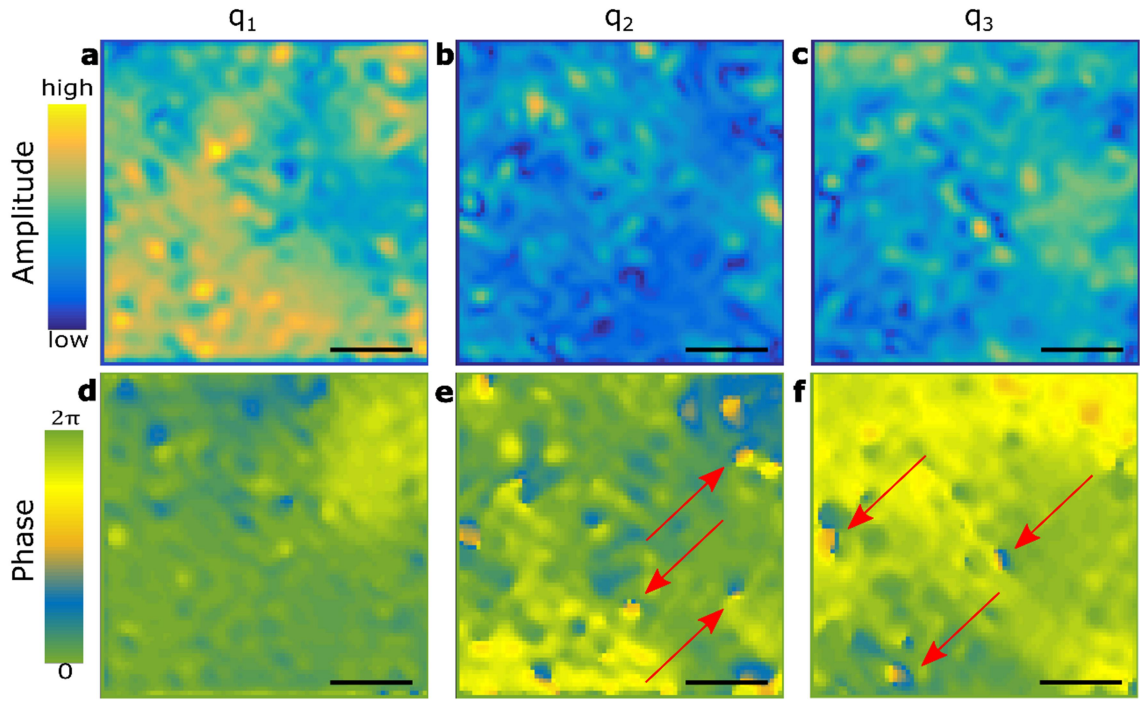

**Fig. 2 | Coexisting charge density waves.** Mapping of the full complex order parameter in 1T-$VSe_2$ for the same surface area as in Fig. 1**a**. **a-c,** The local amplitude of the CDWs in each direction ($q_1$, $q_2$, $q_3$ respectively) and, **d-f,** the corresponding local phase referenced to the lower left corner, whose phase is set to zero. Scale bar: 10 nm. Red arrows are indicating tightly bound vortex-antivortex pairs.

The ability to determine the local amplitude and phase of each CDW enables a much more thorough characterization of the CDW ground state in real space. Beyond explaining the variety of STM imaging contrasts and demonstrating the individual nature of the three CDWs developing in TMDCs, our fitting procedure reveals several intrinsic CDW properties. Of particular interest are singularities in the amplitude and phase, such as domain walls and topological defects (vortices). They are important to understand the interplay of ordered electronic phases[3, 17]. For example, the model analysis discussed here provides a unique contrast mechanism to identify and locate domain walls, which have been proposed to promote superconductivity[18, 19].

We now focus on three selected CDW features to illustrate the augmented experimental phase space accessible by fitting the complex order parameter in real space. The first example is a detailed analysis of $\pi$-phase shift CDW domain walls ($\pi$DWs) developing in TiSe$_2$ when intercalating either Cu [9] or large amounts of Ti [20]. These $\pi$DWs are usually identified by eye, seeking for one atomic row shifts in the characteristic sequence of bright and dark atoms of the $2a \times 2a$ CDW. Such domains can be easily found in Fig 3**a**. Their topographic structure is magnified in Fig. 3**b** and 3**c** corresponding to the red and blue outlined regions in Fig. 3**a**. The real space CDW modulation in these two regions highlighted in the corresponding filtered images (Fig. 3**d** and 3**e**) is perfectly reproduced by our fitting procedure in Fig. 3**f** and 3**g**. Most instructive are the local phases of the three independent CDWs involved in the above fitting procedure (Fig. 3**i-k**). Each phase is essentially constant within any given domain and changes abruptly at the domain wall. An original insight of our analysis is that the domain walls do not necessarily correspond to a $\pi$-shift as shown by the polar histograms in Fig. 3**l-o**. A visual inspection, in the absence of the quantitative phase analysis proposed here, would conclude a $\pi$-shift up to a phase shift close to $\pi/4$, because the same atoms remain the brighter ones within each domain (see Supplementary Information Section 4 and Suppl. Fig. 4). Our holographic analysis further reveals these $\pi$DWs to be one-dimensional line-singularities along which the CDW order parameter is suppressed. The complete fitting data set available in the Supplementary Information indeed shows the amplitude is vanishing at the domain wall (Suppl. Fig. 5**a-c**). Further supporting the collapse of the CDW state at the $\pi$DWs are the low values of the $R^2$ maps along the domain walls (Suppl. Fig. 5**d-f**), indicating that the fitting procedure fails in those locations because there is actually no periodic CDW modulation.

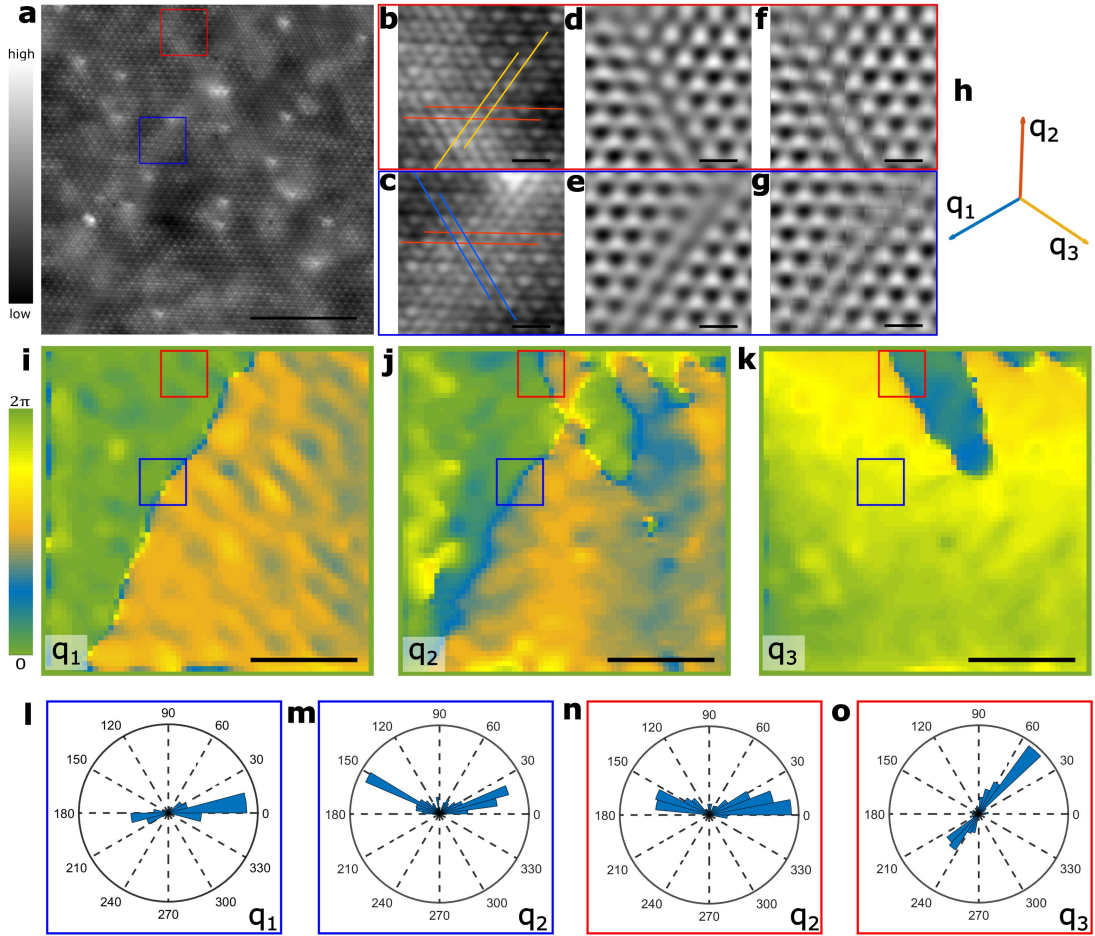

**Fig. 3 | Phase domains of the CDW in 1T-$Cu_{0.02}TiSe_2$ at 1.2 K. a,** High resolution STM micrograph of an in-situ cleaved surface ($V_{bias}$=-200 mV $I_t$= 100 pA). **b** and **c,** selected magnified areas from panel **a** at locations marked by red and blue squares, respectively. Blue ($q_1$), orange ($q_2$) and yellow ($q_3$) parallel lines highlight the shifted CDW modulations for each direction. **d** and **e,** Fourier filtered CDW seen in **b** and **c**. **f** and **g,** are reconstructed images from the mosaics of the fitting procedure. Note the excellent agreement with **d** and **e**. **h,** vectors representing the CDW q-vectors. **i-k,** are the fitted phase images for the three directions ($q_1$, $q_2$, $q_3$). The red and blue squares mark the same areas as in **a**. In the red region, $q_2$ and $q_3$ undergo a phase shift (see panel **b**). This is very well reflected in panels **i-k**: within the red region $q_1$ is homogeneous (no shift) while $q_2$ and $q_3$ undergo an abrupt colour change. Similarly, in the blue region $q_1$ and $q_2$ shift, while $q_3$ does not (see panel **c**). This is very well shown in **i-k**: $q_1$ and $q_2$ show an abrupt colour change in the blue region, but $q_3$ remains uniform. **l-o,** polar histograms of the CDW phase (in degrees) corresponding to: **l,** blue area in **i**; **m,** blue area in **j**; **n,** red area in **j** and **o,** red area in **k**. Scale bars=10 nm in **a**, **i-k**; Scale bars=1 nm in **b-g**.

As a second example, we address discommensurations (DCs), a particular type of defects associated with nearly commensurate CDW (NC-CDW) found for example in 2H-NbSe$_2$ or 2H-TaSe$_2$. Discommensurations were first proposed in the seminal work of McMillan[14] and searched for, e.g in the NC-CDW phase of 1T-TaS$_2$ [21]. DCs are domain walls where the order parameter phase is changing rapidly between small phase locked regions, which allow the CDW system to lower its energy. Our holographic analysis reveals precisely such a phase texture in the CDW images acquired on in-situ cleaved surfaces of 2H-NbSe$_2$ (Fig. 4**a**). Fig. 4**b** is a phase map of one of the three CDWs in the same field of view as Fig. 4**a**. It reveals areas of rather constant phase separated by nanometer-sized regions (domain walls) where the phase is changing rapidly. This staircase structure is best seen in a phase profile (Fig. 4**c**) taken along the red dashed line in Fig. 4**b** which perfectly mimics the prediction by McMillan[14]. The phase-locked region sizes and locations are uncorrelated among the CDW components (Suppl. Fig. 6**d-f**), which confirms their individual nature. We stress that the DC domain walls observed here for the NC-CDW of 2H-NbSe$_2$ are completely different from the πDWs discussed above for 1T-Cu$_{0.02}$TiSe$_2$. The order parameter amplitude remains essentially unchanged along a DC (Suppl. Fig 6**a-c**), whereas it is vanishing along the πDWs (Suppl. Fig. 5**a-c**), suggesting the different nature and origin of these two CDW defects. While πDWs can be readily seen in topographic STM images, DCs can only be detected using our holographic fitting scheme. Note that in order to detect DCs, one has to control the phase wrapping in the fitting procedure. Otherwise, cumulative errors introduce an artificial phase gradient in the phase-locked domains delimited by the DCs, making them essentially undetectable (see Suppl. Fig. 3).

The final example are topological defects (TDs) which are defined as singularities in the order parameter that cannot be annealed without introducing other singularities outside of an arbitrarily small environment of the TDs[22]. They are characterized by a topological invariant, the winding number, which is conserved under any continuous deformation of the order parameter field. They correspond to points in a 2D landscape around which the total phase of the order parameter winds by an integer multiple of $2\pi$. By analogy to vortices in a superconductor, they are called vortices and antivortices depending on the winding polarity. Such defects appear in the phase images of both 1T-VSe$_2$ and 2H-NbSe$_2$. We observe tightly bound vortex-antivortex pairs, as pointed out by red arrows in Fig. 2**e** and 2**f** measured on 1T-VSe$_2$, and well separated vortex-antivortex pairs, as pointed out by white arrows in Suppl. Fig. 6**d-f** measured on 2H-NbSe$_2$. The TD landscape is different for each of the three CDWs in a given STM image, again supporting the conclusion the CDW consists of three individual charge modulations.

Crystalline defects and impurities certainly play role in the amplitude and phase landscape of the CDW phase[15, 16] and in the nucleation near the phase transition[23]. The holographic analysis we propose here reveals more subtle aspects of this interplay. A first revealing insight not available prior to the present experiments is that domain walls and TDs can develop independently in any of the three CDWs as clearly exemplified by the distinct domain patterns of the three CDWs of 1T-$Cu_{0.02}TiSe_2$ in Fig. 3**i-k**. An additional striking manifestation of the coexistence of individual charge orders, highlighted by our holographic analysis, is that a single atom defect can trigger a topological defect in the phase of one CDW, while only producing a smooth phase variation for another one (Fig. 2, Fig. 4**a** and 4**b** and Suppl. Fig. 6). While CDW defects are expected in the presence of structural defects, the holographic analysis discussed here reveals that the CDW order parameter landscape (Fig. 4**e** and 4**f**) can be remarkably inhomogeneous on a 2H-$NbSe_2$ surface with no obvious structural defects (Fig. 4**d**). This is likely the result of discommensurations not readily detectable in the topographic STM images. It may also reflect 3D CDW correlations with defects below the surface not apparent in the STM micrograph. More generally, Fig. 4**d-f** suggests the ability of the CDW to develop a spatial structure essentially independent of the supporting crystalline lattice.

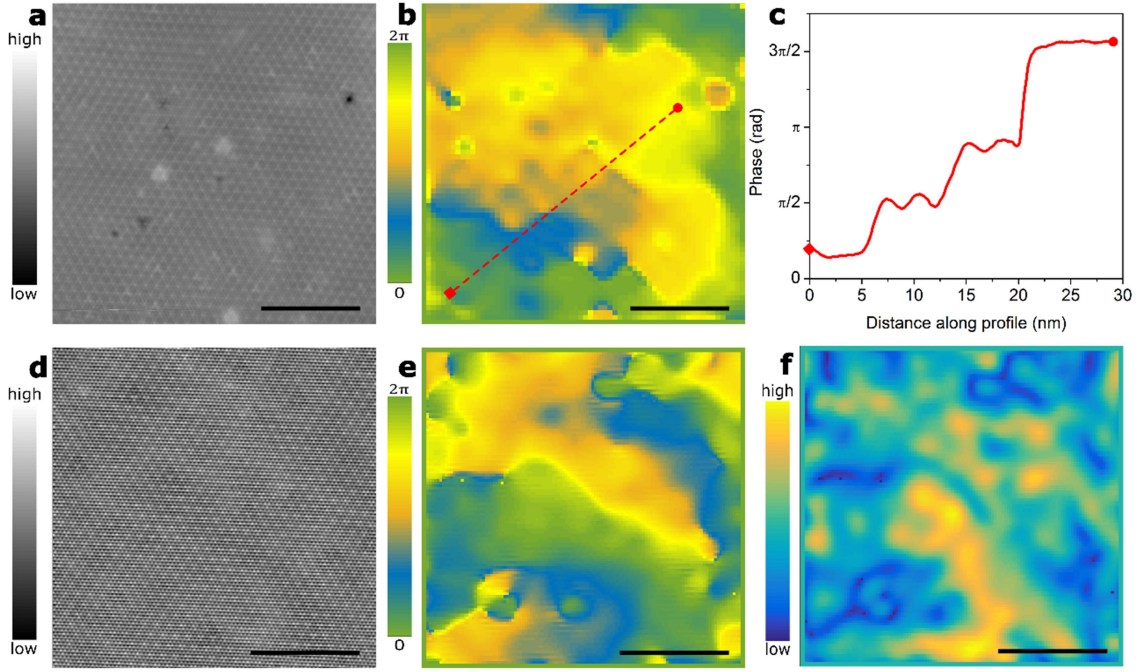

**Fig. 4 | STM imaging of crystalline defects, CDW discommensurations and singularities in 2H-NbSe$_2$. a,** STM micrograph with atomic defects. **b,** Phase map of the corresponding q$_2$ CDW. **c,** Phase profile extracted from the square to the circle along the red dashed line in panel **b**. **d,** STM micrograph of a defect free region on the same sample. **e,** Phase and **f,** amplitude maps of the corresponding q$_2$ CDW. Both STM micrographs were measured at 1.2K on an in-situ cleaved surface. Tunneling parameters: V$_{bias}$=100 mV, I$_t$= 100 pA. Scale bars = 10 nm.

The holographic imaging of the local complex CDW order parameter with a spatial resolution of the order of the CDW wavelength, offers remarkable opportunities to uncover physical details of the CDW state. We find clear evidence for the coexistence of three individual CDWs in transition metal dichalcogenides with a threefold symmetric band structure in the *ab*-plane, each developing a distinct order parameter landscape and a great variety of features. Unambiguous phase images contribute novel insight on the CDW $\pi$DWs reported in 1T-Cu$_{0.02}$TiSe$_2$, where one of the three CDWs does not experience any phase shift, while the phase shift of the remaining two is not necessarily $\pi$. Mastering the phase fitting procedure reveals constant phase domains and provides unprecedented insight into nearly commensurate CDWs, with first real space phase images of individual discommensurations in 2H-NbSe$_2$ predicted long ago by theory. We expect significant contributions to the deeper understanding of charge density waves, including their formation mechanism and their interplay with other correlated electron ground states, from the expanded parameter space accessible through the high-resolution real space mapping of the complex CDW order parameter.

**Methods**

Single crystals of MSe$_2$ (M=Cu$_{0.02}$Ti, Nb, V) were grown via the chemical vapour transport method using Iodine as transport agent and cleaved in-situ at room temperature. STM experiments were done in UHV (base pressure below $2 \cdot 10^{-10}$ mbar) using tips either electrochemically etched from an annealed tungsten wire or mechanically cut from a PtIr wire with similar results. Tips were conditioned in-situ on cleaned and reconstructed Au(111) or Ag(111) single crystals. The bias voltage was applied to the sample. The numerical analysis has been performed in Matlab and Mathematica environments. The complete fitting procedure is detailed in Supplementary Information Section 2.

**Acknowledgement**

We thank A. Guipet and G. Manfrini for their technical support in the STM laboratories. This project was supported by Div.2 and Sinergia of the Swiss National Science Foundation.

**Author contribution**

C.R. designed the experiment. Á.P. and A.S. conceived the model and performed the holographic data analysis. C.B. and E.G. synthesized the bulk crystals. Á.P., A.S. and M.S. performed the STM measurements. Á.P., A.S. and C.R. wrote the paper. All authors contributed to the scientific discussions and manuscript revisions.

**Competing financial interests**

The authors declare no competing financial interests.